\documentclass{article}

\usepackage{spconf}
\usepackage{amsmath}
\usepackage{bm}
\usepackage[nobreak,nocompress]{cite}
\usepackage{graphicx}
\usepackage{hyperref}
\usepackage{siunitx}
\usepackage{multirow}
\usepackage{url}
\usepackage{color}
\usepackage{fancyhdr}

\DeclareMathOperator*{\argmax}{arg\,max}
\fancypagestyle{IEEEcopyright}
{
  \fancyhf{}
  
  \fancyfoot[L]{\footnotesize © 2026 IEEE. Personal use of this material is permitted. Permission from IEEE must be obtained for all other uses, in any current or future media, including reprinting/republishing this material for advertising or promotional purposes, creating new collective works, for resale or redistribution to servers or lists, or reuse of any copyrighted component of this work in other works.}
}

\title{Audio Effect Estimation with DNN-Based Prediction and Search Algorithm}

\twoauthors
{Youichi Okita}
{
    Graduate School of Science and Technology\\
    Kwansei Gakuin University\\
    Sanda, Japan\\
    {\normalsize \href{mailto:okitayouichi2001@gmail.com}{\nolinkurl{okitayouichi2001@gmail.com}}}
}
{Haruhiro Katayose}
{
    School of Engineering\\
    Kwansei Gakuin University\\
    Sanda, Japan\\
    {\normalsize \href{mailto:katayose@kwansei.ac.jp}{\nolinkurl{katayose@kwansei.ac.jp}}}
}

\begin{document}

\ninept

\maketitle
\thispagestyle{IEEEcopyright}

\begin{abstract}
Audio effects play an essential role in sound design.
This research addresses the task of audio effect estimation, which aims to estimate the configuration of applied effects from a wet signal.
Existing approaches to this problem can be categorized into predictive approaches, which use models pre-trained in a data-driven manner, and search-based approaches, which are based on wet signal reconstruction.
In this study, we propose a novel approach that integrates these approaches: first, DNNs predict the dry signal and effect configuration, and then a search is performed based on wet signal reconstruction using these predictions.
By estimating the dry signal in the prediction stage, it becomes possible to complement or improve the predictions using reconstruction similarity as an objective function.
The experimental evaluation showed that methods based on the proposed approach outperformed the method solely based on the predictive approach.
Furthermore, the findings suggest that the task division of predicting the effect type combination followed by the search-based estimation of order and parameters was the most effective across various metrics.
\end{abstract}


\section{Introduction}
Audio effects play an essential role in the sound design of audio content such as music and speech \cite{wilmering2020history}, and have been studied from various perspectives \cite{comunita2024afxresearch}.
Audio effect estimation is the task of estimating the configuration of applied effects from a wet signal, an audio signal after effects have been applied.
An effect configuration consists of a type, which is a categorical label broadly classifying the effect, and its corresponding control parameters.
Moreover, effects are often used as an audio effect chain, where multiple effects are applied in cascade.
Conventionally, achieving a desired sound design using such diverse and complex audio effects has required a high level of expertise in both technical and artistic domains.
Therefore, automatic audio effect estimation enables both novice and expert musicians or audio engineers to efficiently learn and reuse the sound design techniques from existing audio content.

\begin{figure}[t]
    \begin{center}
        \includegraphics[scale=0.55]{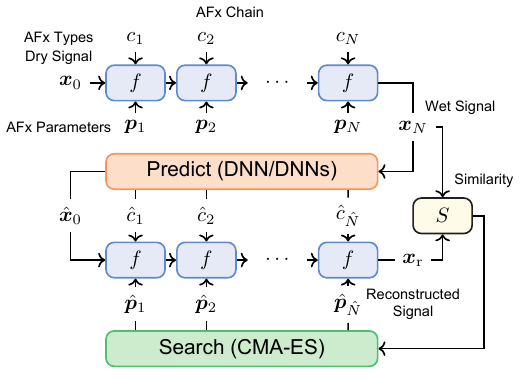}
        \caption{Audio effect estimation with DNN-based prediction and search algorithm.}
        \label{fig:method}
    \end{center}
\end{figure}

Most existing studies on audio effect estimation have taken a predictive approach \cite{jurgens2020recognizing,comunita2021guitar,hinrichs2022convolutional,lee2023blind,guo2023automatic,peladeau2024blind,wada2025hyperbolic}.
In this approach, pre-trained models such as deep neural networks (DNNs) predict the effect configuration on an unseen wet signal.
The models are trained with the error against self-supervised data, consisting of dry signals, wet signals and corresponding effect configurations generated by applying effects to a set of dry signals, as the objective function.
A dry signal is an audio signal with no effects applied.
However, regarding effect chains, most studies have been limited to either parameter regression for fixed combinations and orders of types \cite{hinrichs2022convolutional,peladeau2024blind}, or a multi-label type classification without considering the order \cite{guo2023automatic}.
A predictive approach has also been employed for the task of effect removal, which aims to estimate the dry signal \cite{imort2022distortion,rice2023general,lee2024distortion}.
Moreover, a new task of jointly estimating the dry signal and effect configuration has also been introduced \cite{take2024audio}.
In that study, the model predicts the effect configuration of the last-applied single effect in a chain and the bypass signal, which is the audio signal before the single effect was applied.
By iteratively applying this model to its output bypass signal, the entire chain was estimated.
This made it possible to estimate the configuration of entire effect chain of unknown length and type combination, but the accumulation of errors due to iterative inference remains a challenge.

On the other hand, a search-based approach has also been employed.
In this approach, the effect configuration is optimized using a reconstruction similarity, the similarity between the target wet signal and a reconstructed wet signal generated by applying effects to the dry signal, as an objective function.
This approach has been taken in style transfer for effects \cite{steinmetz2024st,yu2025improving} and mastering \cite{koo2025ito}.
A method using wet signal reconstruction error for training a model in a predictive approach also exists \cite{peladeau2024blind}, but it is limited to differentiable effects \cite{engel2020ddsp}, and the reconstruction similarity is not utilized at inference time.

In this study, as shown in Fig.~\ref{fig:method}, we propose a novel approach that integrates predictive and search-based approaches.
First, a DNN or DNNs predict the dry signal and all or part of the effect configuration, and then a search is performed based on reconstruction similarity using these predictions.
Estimating the dry signal in the prediction stage makes it possible to evaluate the reconstruction similarity, which is then used as an objective function in the search stage to complement or improve the predictions.
When experienced musicians or audio engineers imitate an existing sound design, they typically first make a rough prediction of the effects based on their empirical knowledge and intuition, and then explore and adjust the order and parameters of the effects while monitoring the reconstructed wet signal.
Our proposed approach follows a similar process.

We evaluate our methods on wet signals generated by applying effect chains to musical excerpts played on the guitar.
We compare multiple settings of task division between prediction and search stage in the evaluation.
We also provide examples of the results online\footnote{\url{https://okitayouichi.github.io/afx-pred-sch-demo/}}.

\section{Problem Formulation}
We formulate the audio effect chain estimation task addressed in this study.
First, the application of a single audio effect to an audio signal can be expressed as $\bm{x}_1=f_{c,\bm{p}}\left(\bm{x}_0\right)$, where $\bm{x}_0$ is the dry signal, $\bm{x}_1$ is the wet signal, $f_{c,\bm{p}}$ is the effect, $c$ is its type, and $\bm{p}$ are the parameters corresponding to $c$.
Next, we consider an audio effect chain.
The application of the chain of length $N$ can be expressed as
\begin{equation}
    \label{eq:fx_apply_chain}
    \bm{x}_N = f_{c_N,\bm{p}_N} \circ \cdots \circ f_{c_1,\bm{p}_1}\left(\bm{x}_0\right) = F_{C,P}\left(\bm{x}_0\right).
\end{equation}
The dry signal is $\bm{x}_0$, and the wet signal is $\bm{x}_N$.
We have introduced the notation for the ordered sequences of types and parameters, $C=\left(c_1,\cdots,c_N\right)$ and $P=\left(\bm{p}_1,\cdots,\bm{p}_N\right)$, respectively, and $F_{C,P}$ to represent the entire effect chain set by them.

Audio effect chain estimation can be expressed as
\begin{equation}
    \label{eq:fx_chain_est}
    \left(\hat{C},\hat{P},\hat{\bm{x}}_0\right)=E\left(\bm{x}_N\right),
\end{equation}
where $E$ is the estimator, which estimates the sequence of the effect configuration $\left(\hat{C},\hat{P}\right)$ and the dry signal $\hat{\bm{x}}_0$ from the wet signal $\bm{x}_N$.
The chain length $\hat{N}$ is also estimated through this process.

\section{Proposed Methods}
In this study, we propose a two-stage approach for the task represented by Eq.~\eqref{eq:fx_chain_est}, as shown in Fig.~\ref{fig:method}, which consists of DNN-based prediction and search based on wet signal reconstruction.
In a two-stage approach, the division of tasks between them is an essential design choice.
Therefore, we define three settings of task division for comparison.

\subsection{Prediction with Deep Neural Networks}

\subsubsection{Problem Formulation and Prediction Methods}
We define three settings for the prediction stage.
The search stage then handles the remaining estimation tasks.

\noindent \textbf{Dry-Type-Direct.}~
The DNN $g_1$, on the entire chain, predicts the unordered combination of effect types and the dry signal directly:
\begin{equation}
    \label{eq:dry_type_direct}
    \left(\left\{\hat{c}_1,\cdots,\hat{c}_{\hat{N}}\right\}, \hat{\bm{x}}_0\right)=g_1\left(\bm{x}_N\right).
\end{equation}
The type prediction is a multi-label classification.

\noindent \textbf{Bypass-Type-Iter.}~
The DNN $g_2$, on the last-applied single effect in the chain, predicts the effect type and the bypass signal:
\begin{equation}
    \label{eq:dry_type_iter}
    \left(\hat{c}_n,\hat{\bm{x}}_{n-1}\right)=g_2\left(\bm{x}_{n}\right).
\end{equation}
The type prediction is a single-label classification.
In this task, the empty chain is also a target for estimation and is predicted as a special class ``None''.
This model predicts by iteratively feeding its output $\hat{\bm{x}}_{n-1}$ back into $g_2$ itself until $\hat{c}_n=\mathrm{None}$.
By starting this iterative inference from the wet signal, it predicts the ordered sequences of audio signals $\left(\hat{\bm{x}}_0,\cdots,\hat{\bm{x}}_{\hat{N}-1}\right)$ and effect types $\hat{C}$.

\noindent \textbf{Bypass-Config-Iter.}~
The DNN $g_3$, on the last-applied single effect in the chain, predicts the effect configuration and the bypass signal:
\begin{equation}
    \label{eq:dry_config_iter}
    \left(\hat{c}_n,\hat{\bm{p}}_n,\hat{\bm{x}}_{n-1}\right)=g_3\left(\bm{x}_n\right).
\end{equation}
As with Bypass-Type-Iter, the type prediction is a single-label classification and also predicts the special class for an empty chain.
By starting the same iterative inference as Bypass-Type-Iter, it predicts the sequences of audio signals $\left(\hat{\bm{x}}_0,\cdots,\hat{\bm{x}}_{\hat{N}-1}\right)$ and effect configurations $\left(\hat{C},\hat{P}\right)$.
This task is equivalent to that addressed by SunAFXiNet \cite{take2024audio}.

Hereafter, Bypass-Type-Iter and Bypass-Config-Iter may also be written collectively as \textbf{Bypass-*-Iter}.

\subsubsection{Network Architecture}

\begin{figure}[t]
    \begin{center}
    \includegraphics[scale=0.55]{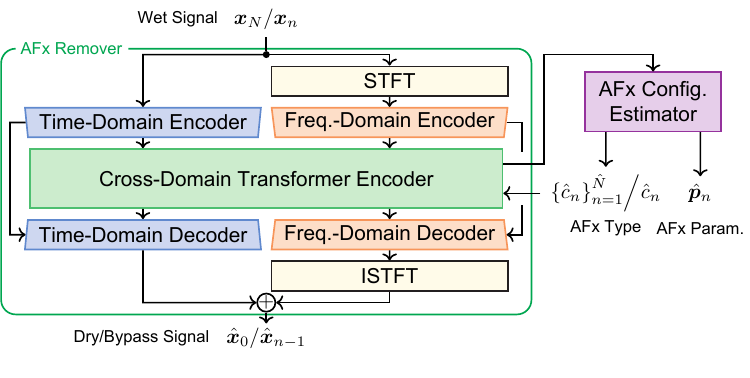}
    \caption{Architecture of the prediction model.}
    \label{fig:model}   
    \end{center}
\end{figure}

\begin{figure}[t]
    \begin{center}
    \includegraphics[scale=0.55]{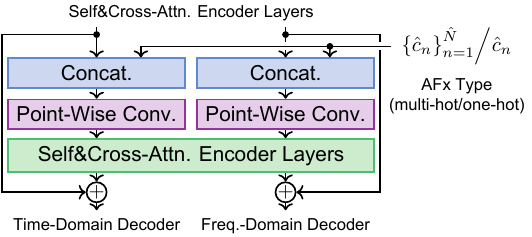}
    \caption{Conditioning with effect type in the cross-domain encoder.}
    \label{fig:cde} 
    \end{center}
\end{figure}

The DNN architecture follows SunAFXiNet \cite{take2024audio}.
Unless otherwise specified, the model's hyperparameters are also adopted from that.
As shown in Fig.~\ref{fig:model}, this model consists of an effect remover that processes signals in the time and frequency domains with U-Net-based networks, and an effect configuration estimator that branches from the cross-domain encoder at its bottleneck.
The effect remover, similar to previous studies \cite{hinrichs2024blind,take2024audio}, is based on Hybrid Transformer Demucs \cite{rouard2023hybrid}.

The cross-domain encoder consists of normalization and positional encoding, followed by alternating five layers of self-attention and cross-attention encoders.
The effect configuration estimator branches off after the third layer, and the conditioning by the effect type, one of its outputs, is performed before the fourth layer.

The effect configuration estimator consists of, for each domain, three layers of convolutional blocks, global pooling along the time axis, concatenation of signals from both domains in the channel axis, and three layers of fully connected blocks for type classification.
Each convolutional block preserves the number of channels, while the fully connected blocks halve it.
For Bypass-Config-Iter, an additional parameter regression branch consisting of three layers of fully connected blocks is included.
This branch outputs parameters of all possible types regardless of the type classification, allowing for training and evaluation irrespective of whether the classification is correct.
In the final layer of the type classification branch of the Dry-Type-Direct, sigmoid is applied, and the output is converted to a multi-hot representation with a threshold of 0.5 for conditioning.
In Bypass-*-Iter, softmax is applied, and it is converted to a one-hot representation by argmax operation for conditioning.
As illustrated in Fig.~\ref{fig:cde}, the conditioning is performed by duplicating the effect type representation across all time steps, concatenating it with the signal of each domain along the channel axis, and then restoring the original number of channels using a point-wise convolution.

\subsubsection{Training Process}
The models are trained in two stages, following prior work \cite{take2024audio}.

In the first stage, only the effect remover is trained.
The loss functions for Dry-Type-Direct and Bypass-*-Iter are, respectively,
\begin{equation}
    \label{eq:loss11}
    L_{11}=L_{\mathrm{mae}}\left(\hat{\bm{x}}_{0},\bm{x}_{0}\right) + \alpha L_{\mathrm{mrstft}}\left(\hat{\bm{x}}_{0},\bm{x}_{0}\right),
\end{equation}
\begin{equation}
    \label{eq:loss123}
    L_{12}=L_{\mathrm{mae}}\left(\hat{\bm{x}}_{n-1},\bm{x}_{n-1}\right) + \alpha L_{\mathrm{mrstft}}\left(\hat{\bm{x}}_{n-1},\bm{x}_{n-1}\right),
\end{equation}
where $L_{\mathrm{mae}}, L_{\mathrm{mrstft}}, \alpha$ are the Mean Absolute Error (MAE), Multi-Resolution STFT loss (MR-STFT) \cite{yamamoto2020parallel,steinmetz2020auraloss} and the weight of the MR-STFT, respectively.
In the training of Bypass-*-Iter, the input signal itself is used as the ground truth bypass signal when the target is an empty chain.
Also, the ground-truth effect type is used for conditioning.
The effect remover's parameters are identical for Bypass-*-Iter.

In the second stage, the parameters of the effect remover are frozen, and only the effect configuration estimator is trained.
The loss functions for Dry-Type-Direct, Bypass-Type-Iter, and Bypass-Config-Iter are, respectively,
\begin{equation}
    L_{21}=L_{\mathrm{bce}}\left(\left\{\hat{c}_1,\cdots,\hat{c}_{\hat{N}}\right\},\left\{{c}_1,\cdots,{c}_{N}\right\}\right),
\end{equation}
\begin{equation}
    L_{22}=L_{\mathrm{ce}}\left(\hat{c}_n,c_n\right),
\end{equation}
\begin{equation}
    L_{23}=L_{\mathrm{ce}}\left(\hat{c}_n,c_n\right)+L_{\mathrm{mse}}\left(\hat{\bm{p}}_n,\bm{p}_n\right),
\end{equation}
where $L_{\mathrm{bce}}, L_{\mathrm{ce}},$ and $L_{\mathrm{mse}}$ are Binary Cross-Entropy, Cross-Entropy, and Mean Squared Error, respectively.
Also, during the training of parameter regression, only the parameters corresponding to the ground-truth type are used.

\subsection{Search Based on Wet Signal Reconstruction}

\subsubsection{Problem Formulation}
In the search stage, the core problem is an optimization for the effect parameters that maximizes the similarity between the reconstructed wet signal $\bm{x}_{\mathrm{r}}=F_{\hat{C},P}\left(\hat{\bm{x}}_0\right)$ and the given wet signal:
\begin{equation}
    \label{eq:reconst}
    \hat{P} = \argmax_P{S\left(F_{\hat{C},P}\left(\hat{\bm{x}}_0\right),\bm{x}_N\right)},
\end{equation}
where $S$ is the similarity between audio signals, and in this study, we use the Scale-Invariant Signal-to-Distortion Ratio (SI-SDR) \cite{le2019sdr}.

The predicted value in the preceding stage is used as $\hat{\bm{x}}_0$.
For Dry-Type-Direct, which does not predict the order of $\hat{C}$, the search is performed in two stages.
First, the search is performed for all permutations of the predicted combination of types, and the one with the best similarity is taken as the estimated type sequence.
Then, an additional search is performed on that sequence for further refinement.
In this case, the $\hat{P}$ obtained in the first stage is used as the initial solution in the second stage.
Also, for Bypass-Config-Iter, $\hat{P}$ obtained in the prediction stage is used as the initial solution.

\subsubsection{Search Algorithm}
The implementation of the effect chain $F_{\hat{C},P}$ that appears in the objective function $S$ is often unknown.
Therefore, this problem requires black-box optimization.
As the optimization algorithm, we primarily employed the Covariance Matrix Adaptation Evolution Strategy (CMA-ES) \cite{hansen1996adapting}, following related work \cite{steinmetz2024st}.
Since CMA-ES is ineffective when the number of search parameters is 1, we employed the Tree-structured Parzen Estimator \cite{bergstra2011algorithms} instead.

\section{Experimental Evaluation}

\subsection{Dataset}

\begin{table}[t]
    \caption{Pedalboard effects used for dataset generation.} 
    \label{tab:fx}
    \hbox to\hsize{\hfil
    \begin{tabular}{ccc}
    \hline
    Type & Variable Parameter & Range \\
    \hline
    \multirow{3}{*}{Chorus} 
    & depth & 0.1--0.3 \\
    & feedback & 0.0--0.5 \\
    & mix & 0.3--0.7 \\
    \hline
    Distortion & drive\_db & 10.0--20.0 \\
    \hline
    \multirow{3}{*}{Reverb}
    & room\_size & 0.1--0.7 \\
    & damping & 0.1--0.9 \\
    & wet\_level & 0.1--0.4 \\
    \hline
    \end{tabular}\hfil}
\end{table}

First, we collected dry signals from existing datasets.
We extracted a total of 2231 non-overlapping chunks of $\SI{10.0}{\second}$ from musical excerpts played on the guitar without effects applied, from IDMT-SMT-Guitar \cite{kehling2014automatic}, GuitarSet \cite{xi2018guitarset}, EGDB \cite{chen2022towards}, and Guitar-TECHS \cite{pedroza2025guitar}.
The number of channels was unified to 1, the sampling frequency to $\SI{44.1}{\kilo\hertz}$, and the level was normalized to Root Mean Square (RMS) of 0.1.

Next, we generated wet signals by applying effects to dry signals.
We used the pedalboard\footnote{\url{https://github.com/spotify/pedalboard}} library for the effects.
Considering practical sound design and estimability, we selected the types, variable parameters, and their ranges as shown in Tab.~\ref{tab:fx}.
From these 3 types, we formed effect chains by selecting each type at most once, and including the intermediate signals, we generated $\sum_{n=1}^{3}{n{}_3 P_{n}}=33$ wet signals per dry signal.
The parameters were randomly determined within the ranges, and for those not listed, the library's default values were used.
These were normalized to $\left[0,1\right]$ during training and evaluation.
We normalize the audio level to an RMS of 0.1 after each effect in the chain to focus on the core characteristics of the effects rather than on level changes.
Furthermore, to ensure the reproducibility of handling audio via files, values outside the range $\left[-1.0,1.0\right]$ were clipped.

As a result, the total duration of input signal entries was $2231\times33\times\SI{10.0}{\second}=\SI{205}{\hour}$.
For the training and evaluation of Bypass-*-Iter, which also predict empty chains, we add empty chain entries.
The number of those entries was equal to the number of entries where another effect was last-applied, which was $33/3=11$ per dry signal.

\subsection{Experimental Setup}
The dataset was split into 80\% for training, 15\% for validation, and 5\% for evaluation, ensuring no overlap of dry signal tracks.

In the training, the weight of the MR-STFT in Eq.~\eqref{eq:loss11} and \eqref{eq:loss123} was set to $\alpha=0.01$.
We employed the AdamW optimizer \cite{loshchilov2019decoupled} with learning rates of $1\times10^{-4},1\times10^{-5}$ for the first and second stages, respectively.
The weight decay was set to $1\times10^{-2}$.
We also employed gradient clipping with a value of 5.0.
The batch size was set to 64 and the number of epochs was set to 170 and 50 for the first and second stages, respectively.
The validation metric was SI-SDR of the dry or bypass signal for the first stage, and Macro $F_1$ Score of types for the second stage.
The model state with the best validation metric value was used for evaluation.

In the search stage, the total number of trials was set to $M=\lfloor M_0 d^r \rfloor$, where $d$ is the search dimension.
Based on preliminary experiments, we set $M_0=5$ for the first stage for Dry-Type-Direct, $M_0=20$ for the second stage for Dry-Type-Direct and the Bypass-*-Iter, and $r=1.5$ for all cases.
We employ the implementation in Optuna \cite{akiba2019optuna} for search algorithms, following its default settings unless otherwise specified.

In the training, empty chain entries were used solely to enable the models to learn stopping condition of the iterative inference in Bypass-*-Iter.
Therefore, in the evaluation of effect chain removal (Tab.~\ref{tab:dry}) and wet signal reconstruction (Tab.~\ref{tab:reconst}), ground-truth and predicted empty chains were excluded to avoid an overestimation.
The stopping condition for the iterative inference was that the model predicted the special class ``None'' or the chain length reached $\hat{N}=3$.

\subsection{Evaluation Results}

\subsubsection{Audio Effect Configuration Estimation}

\begin{table}[t]
    \caption{Evaluation results for single audio effect type classification.} 
    \label{tab:type_single}
    \hbox to\hsize{\hfil
    \begin{tabular}{cc}\hline
    Method & Macro $F_1$ \\\hline
    Bypass-Type-Iter & \textbf{0.919} \\
    Bypass-Config-Iter & 0.917 \\\hline 
    \end{tabular}\hfil}
\end{table}

First, for the models that perform iterative inference in the prediction stage, we evaluated the type classification of the last-applied single effect in the chain.
Tab.~\ref{tab:type_single} shows the results using the Macro $F_1$ Score.

\begin{table}[t]
    \caption{Evaluation results for audio effect chain type classification.} 
    \label{tab:type_chain}
    \hbox to\hsize{\hfil
    \begin{tabular}{cccc}\hline
    Method & Macro $F_1$ & LD & EMA \\\hline
    Dry-Type-Direct + Search & \textbf{0.958} & \textbf{0.313} & \textbf{0.774} \\
    Bypass-Type-Iter & 0.949 &  0.369 & 0.723 \\
    Bypass-Config-Iter & 0.942 & 0.408 & 0.702 \\\hline
    \end{tabular}\hfil}
\end{table}

Next, we evaluated the type classification of the entire effect chain.
Tab.~\ref{tab:type_chain} shows the results using Macro $F_1$ Score, which does not consider order, and Levenshtein Distance (LD) and Exact Match Accuracy (EMA), which do consider order.
LD can also evaluate the partial correctness of a sequence.
The Dry-Type-Direct with the search achieved the best performance across all metrics.
The performance degradation on the entire chain classification using Bypass-*-Iter is likely attributable to error accumulation.

For Bypass-Config-Iter, which estimates effect parameters in the prediction stage, the evaluation of the last-applied effect resulted in an MAE of 0.0885 for the parameters.

\subsubsection{Audio Effect Removal}

\begin{table}[t]
    \caption{Evaluation results for single audio effect removal.} 
    \label{tab:bypass}
    \hbox to\hsize{\hfil
    \begin{tabular}{ccc}\hline
    Method & SI-SDR & MR-STFT \\\hline
    Bypass-Type-Iter & \textbf{26.32} & \textbf{0.690} \\
    Bypass-Config-Iter & 26.30 & 0.691 \\\hline 
    \end{tabular}\hfil}
\end{table}

First, for the models that perform iterative inference in the prediction stage, we evaluated the  bypass signal estimation after removing the last-applied single effect in the chain.
Tab.~\ref{tab:bypass} shows the results using SI-SDR and MR-STFT.

\begin{table}[t]
    \caption{Evaluation results for audio effect chain removal.} 
    \label{tab:dry}
    \hbox to\hsize{\hfil
    \begin{tabular}{ccc}\hline
    Method & SI-SDR & MR-STFT \\\hline
    Dry-Type-Direct & 13.96 & \textbf{0.813} \\
    Bypass-Type-Iter & \textbf{14.95} & 0.898 \\
    Bypass-Config-Iter & 14.88 & 0.902 \\\hline
    \end{tabular}\hfil}
\end{table}

Next, we evaluated the dry signal estimation after removing the entire effect chain.
The results are shown in Tab.~\ref{tab:dry}.
Bypass-Type-Iter performed best on SI-SDR, while Dry-Type-Direct was best on MR-STFT.

\subsubsection{Wet Signal Reconstruction}

\begin{table}[t]
    \caption{Evaluation results for wet signal reconstruction.} 
    \label{tab:reconst}
    \hbox to\hsize{\hfil
    \begin{tabular}{ccc}\hline
    Method & SI-SDR & MR-STFT \\\hline
    Bypass-Config-Iter (Baseline) & 18.18 & 0.465 \\
    Dry-Type-Direct + Search & \textbf{23.07} & \textbf{0.340} \\
    Bypass-Type-Iter + Search & 22.68 & 0.361 \\
    Bypass-Config-Iter + Search & 22.64 & 0.366 \\\hline
    \end{tabular}\hfil}
\end{table}

We evaluated the reconstruction of the wet signal after the search stage.
To evaluate the performance of effect configuration estimation independently of the performance of effect removal, we performed the reconstruction using ground-truth dry signals.
The results are shown in Tab.~\ref{tab:reconst}.
We also show the results of Bypass-Config-Iter without the search as a baseline.
All methods combined with the search outperformed the baseline, demonstrating the effectiveness of the proposed approach.
Among the task division strategies for the prediction and search stages, Dry-Type-Direct with the search achieved the best performance on both metrics.

\section{Conclusion}
This study proposed an approach for audio effect estimation that integrates predictive and search-based approaches.
The experimental evaluation showed that methods based on the proposed approach outperformed the method solely based on the predictive approach.
Furthermore, the findings suggest that the task division of predicting the effect type combination followed by the search-based estimation of order and parameters was the most effective across various metrics.

Finally, we mention the limitations of this study: the restricted diversity of the effects handled.
For example, the rate of the Chorus is crucial for its characteristics but we observed that it has a steep, hard-to-optimize landscape in the similarity space, so we fixed its value to a constant.
Furthermore, other effect types not covered in this study and longer effect chains are also used in practice.
Addressing these limitations by exploring effective methods for more diverse effects will be future work.

\newpage
\bibliographystyle{IEEEbib}
\bibliography{afx_estimate_short}

\end{document}